\documentclass{aa}   
\usepackage[varg]{txfonts}
\usepackage[normalem]{ulem}

\usepackage{graphicx, natbib}
\bibpunct{(}{)}{;}{a}{}{,}

\usepackage{amssymb}
\begin{document}
%\begin{frontmatter}
\title{Assessment of detectability of neutral interstellar deuterium by IBEX observations}
\author{M.A. Kubiak \inst{1} \and 
M. Bzowski \inst{1} \and 
J.M. Sok\'o\l \inst{1} \and 
E. M{\"o}bius \inst{2} \and
D.F. Rodr{\'i}guez \inst{3} \and
P. Wurz \inst{3} \and
D.J. McComas \inst{4}}
\offprints{M.Bzowski (\email{bzowski@cbk.waw.pl})}
\institute{Space Research Centre, Polish Academy of Sciences, Bartycka 18A, 00-716 Warsaw, Poland
      \email{mkubiak@cbk.waw.pl}   \and 
Space Science Center and Department of Physics, University of New Hampshire \and
Physics Institute, University of Bern, 3012 Bern, Switzerland \and
Southwest Research Institute, P.O. Drawer 28510, San Antonio, TX  78228, USA
      }
\date{}

\abstract{The abundance of deuterium in the interstellar gas in front of the Sun gives insight into the processes of filtration of neutral interstellar species through the heliospheric interface and potentially into the chemical evolution of the Galactic gas .}
{We investigate the possibility of detection of neutral interstellar deuterium at 1~AU from the Sun by direct sampling by the Interstellar Boundary Explorer (IBEX).} 
{Using both previous and the most recent determinations of the flow parameters of neutral gas in the Local Interstellar Cloud and an observation-based model of solar radiation pressure and ionization in the heliosphere, we simulate the flux of neutral interstellar D at IBEX for the actual measurement conditions. We assess the number of interstellar D atom counts expected during the first three years of IBEX operation. We also simulate observations expected during an epoch of high solar activity. In addition, we calculate the expected counts of D atoms from the thin terrestrial water layer, sputtered from the IBEX-Lo conversion surface by neutral interstellar He atoms.} 
{Most D counts registered by IBEX-Lo are expected to originate from the water layer, exceeding the interstellar signal by 2 orders of magnitude. However, the sputtering should stop once the Earth leaves the portion of orbit traversed by interstellar He atoms. We identify seasons during the year when mostly the genuine interstellar D atoms are expected in the signal. During the first 3 years of IBEX operations about 2 detectable interstellar D atoms are expected. This number is comparable with the expected number of sputtered D atoms registered during the same time intervals.} 
{The most favorable conditions for the detection occur during low solar activity, in an interval including March and April each year. The detection chances could be improved by extending the instrument duty cycle, e.g., by making observations in the special deuterium mode of IBEX-Lo.}


 
%\authorrunning{Kubiak et al.}
%\titlerunning{Detectability of interstellar D by IBEX}

\maketitle
%\end{frontmatter}
\section{Introduction}

The abundance of neutral interstellar deuterium (NIS~D) in the Local Interstellar Medium has up to now been investigated mostly using EUV spectroscopy methods \citep{hebrard_moos:03a, linsky:07}, even though it was also positively detected in the visual domain \citep{hebrard_etal:00a}. Within $\sim$100~pc from the Sun it was determined to be $\xi_{\mathrm{NISD}}=15.6\pm0.4$~ppm by number \citep{linsky_etal:06a}, while in the Local Interstellar Cloud (LIC), represented by the line of sight towards Sirius, it is $\xi_{\mathrm{NISD}}=16 \pm 4$~ppm \citep{hebrard_etal:99a}. The unique capability of the Interstellar Boundary Explorer (IBEX) \citep{mccomas_etal:09a} to directly detect neutral interstellar H \citep{mobius_etal:09b, saul_etal:12a}, He \citep{mobius_etal:09b, mobius_etal:12a, bzowski_etal:12a}, Ne and O \citep{mobius_etal:09b, bochsler_etal:12a} raises the question whether the IBEX-Lo neutral atom camera \citep{fuselier_etal:09b} could potentially also detect NIS~D.  

IBEX-Lo is a time-of-flight mass spectrometer that detects and analyzes negative ions originating from the impact of neutral atoms on a specially designed conversion surface \citep{wurz_etal:06a}. Neutral interstellar atoms are detected when they bounce off the conversion surface, capturing an electron and thus turning into a negative ion. Owing to their unique mass-to-charge ratio and the known energy of the interstellar atoms, the NIS~D signature should be possible to identify in the IBEX-Lo data. 

The distribution of NIS D in the heliosphere has drawn relatively little attention up to now. For many years, the only paper to touch the subject was by \citet{fahr:79}, who indicated that the solar resonant radiation pressure for D should be approximately half of the pressure acting on H because of the atomic mass difference. Before the IBEX launch, the detectability of NIS~D had been addressed by \citet{tarnopolski:07} and  \citet{tarnopolski_bzowski:08a}. They suggested that potentially, the beam of NIS~D atoms should be visible for IBEX during two time intervals each year: August through November and January through March. However, the detection sensitivity of IBEX-Lo strongly drops with decreasing particle energy, and thus only the January -- March interval may be promising. During this interval, the energy of the atoms that enter the sensor is increased because the Earth along with IBEX is moving into the incoming flow. The expected total deuterium flux was assessed as 0.015~cm$^{-2}$ s$^{-1}$.

\begin{figure}
\resizebox{\hsize}{!}{\includegraphics[angle=-90]{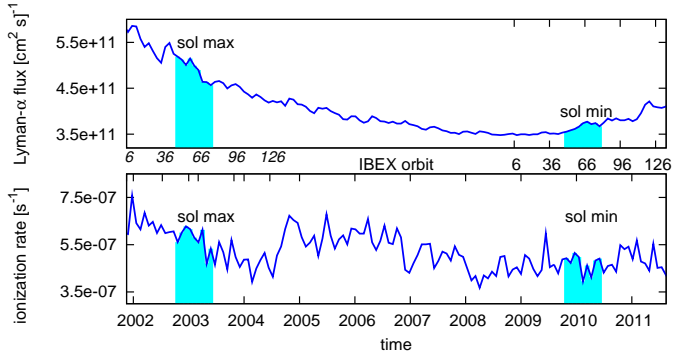}}
\caption{Carrington rotation averaged composite Lyman-$\alpha$ flux \citep[][upper panel]{woods_etal:00} and total ionization rate of D and H (lower panel) since the beginning of the descending phase of solar activity in 2002 until present. The cyan areas mark the solar maximum interval (left) and the solar minimum interval (right) used in the calculations. The placement of IBEX orbits in time relative to the solar activity phase is shown between the panels; italics mark the artificial orbits which are shifted to the beginning of 2002.}
\label{fig:jon}
\end{figure}

\begin{table*}
\caption{Flow parameters of H, D, and He at TS used in the simulations}
\label{tab:isParams}
\centering
\begin{tabular}{ l c c r c l}
\hline \hline
population&$\lambda \, [\degr]$&$\phi\, [\degr]$&$T$ [K]&$V$ [km  s$^{-1}$]&$n_{\mathrm 0}$~[cm$^{-3}$]\\
\hline 
primary~$_{\mathrm{H, D}}\,^{(1)}$& 255.4&5.31&6020 &28.512&0.0312$^{(2)}$\\

secondary~$_{\mathrm{H, D}}\,^{(1)}$& 255.4&5.31&16300 &18.744&0.0542$^{(2)}$\\

NISHe~$_{\mathrm{He, D, O}}\,^{(3)}$&259.2&5.12&6165 &22.756&0.0150$^{(4)}$ \\
\hline 
\end{tabular}
\tablefoot{ The densities of the primary and secondary populations are listed for H, the density labeled as  NIS~He population is for He. The densities of D populations were adopted as for H, multiplied by the interstellar abundance $\xi_{\mathrm{NISD}}=15.6$~ppm (see text). The density of O at TS was adopted $5\times10^{-3}$~cm$^{-3}$ after \citet{slavin_frisch:07a}.}
\tablebib{(1)~\citet{izmodenov_etal:03b}; (2) \citet{bzowski_etal:08a}; (3)~\citet{bzowski_etal:12a}; (4) \citet{witte:04}.}

\end{table*}

In practice, the conversion surface of IBEX-Lo is permanently covered with a thin layer of terrestrial water, whose molecules are sputtered off by the incoming neutral interstellar atoms and dissociated. Some of the sputter products are negative ions, which are registered in IBEX-Lo similar to converted negative ions and, with their energy and composition signature, allow the analysis of neutral interstellar He and Ne \citep{mobius_etal:12a, bochsler_etal:12a}. Since hydrogen in terrestrial water contains $\xi_{\mathrm{terr}} = 150$~ppm deuterium \citep{boehlke_etal:05a}, IBEX-Lo will also detect D$^-$ ions from the water layer, sputtered primarily by NIS~He atoms, thus producing a substantial foreground that competes with NIS~D atoms.

\begin{figure*}[t!]
\centering
{\includegraphics[width=5.2cm,height=\textwidth,angle=-90]{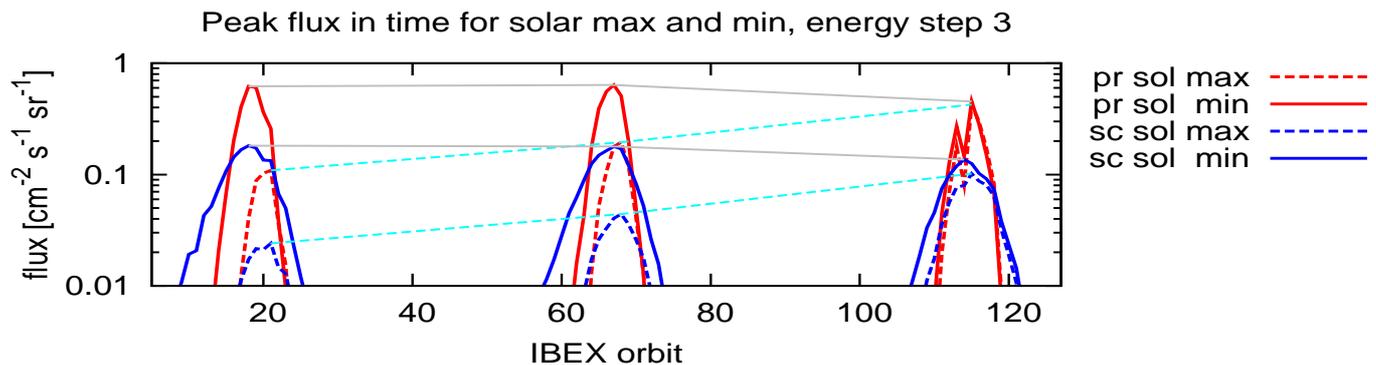}}\\
\caption{Evolution in time of the peak of energy-integrated NIS~D flux at IBEX as a function of IBEX orbit number for solar activity conditions rising from minimum and declining from maximum, calculated for the primary (pr, red) and secondary (sc, blue) populations of NIS~D (for parameters used, see Table \ref{tab:isParams}). Solid lines show the evolution for the solar minimum conditions, broken lines for the solar maximum conditions. Pale blue dashed thin lines show the evolution of the yearly maximum of the signal when the solar activity decreases (the top line is for the primary population, the bottom for the secondary). Gray solid lines show the decrease of the signal maximum with the increase of solar activity from the minimum conditions in 2009. Because of the energy sensitivity threshold, only the peaks around orbits 20, 67, and 115 are accessible to IBEX-Lo. The ``dents'' are due to specifics of the IBEX spin axis pointing.}
\label{fig:3seasons}
\end{figure*}

NIS atoms of various species observed by IBEX, when mapped on the sky, form well defined, oval regions \citep{mobius_etal:09b}. Details of the process of observations of the NIS gas flux by IBEX-Lo were presented by \citet{mobius_etal:12a}. In brief, the spacecraft is spin-stabilized and the optical axis of the detector is perpendicular to the line of sight \citep{hlond_etal:12a}. The axis is fixed during one IBEX orbit and the field of view is a strip $7\degr$-wide, straddling a great circle on the sky nearly perpendicular to the ecliptic plane. The spacecraft records counts of NIS atoms of various species as a function of spin angle. The spin axis is shifted at each new orbit to approximately follow the Sun and thus a different strip on the sky is imaged on each orbit. 

IBEX-Lo is capable to observe NIS atoms with energies from $\sim 10$~eV to $\sim 2$ keV \citep{fuselier_etal:09b}. This interval is divided into eight partly overlapping energy ranges $E_i$, $i = 1, ..., 8$, whose central energies are equal to 14.5, 28.5, 55.5, 107, 207.5, 451, 909, and 1903 eV, respectively, and the widths are proportional to the respective central energies: $\Delta E_i/E_i \simeq 0.7$. These energy ranges, referred to as the IBEX-Lo energy steps, are switched sequentially during the observations.  

The observed count number of atoms of a given species is proportional to the time-integrated flux of NIS atoms in the IBEX-inertial reference frame. The data may be analyzed as a function of spin angle, which closely corresponds to ecliptic latitude, or they may be integrated over spin angle and then the total number of atoms observed on a given orbit regardless of their arrival direction is obtained.

Because now we know the exact time intervals and geometry of the IBEX-Lo interstellar flow observations for three seasons after the IBEX launch at the end of 2008 \citep{mobius_etal:12a, hlond_etal:12a}, we can simulate the expected NIS~D flux as well as the internal D production and thus determine the actual detection conditions. We also consider the NIS~D detectability during solar maximum. We identify the preferred solar activity phases for NIS~D detection in the Earth orbit, specify operational requirements, and suggest some measures to improve NIS~D detection chances in future observations.

This paper explores through modeling the detection conditions of NIS D, but does not discuss how to extract the NIS D component from the deuterium signal registered by IBEX-Lo.  This highly complex task deserves a separate, specially tailored study, which is now in progress (Rodr{\'i}guez, private communication). The paper discusses a potential opportunity to add the technique of in-situ measurements of NIS D to the body of knowledge of interstellar medium. As typical in scientific research, the new technique starts with a demonstration what can be done, what conditions exist and how they affect the practical application of this new technique. The modeling paper will hopefully be followed by a first presentation of the search results, likely with large error bars determined by available observing conditions. With time, based on the first insight, the new technique can be refined (possibly using a better optimized instrument) to provide statistics large enough to draw quantitative conclusions on the density of NIS D in front of the heliosphere and its abundance relative to other NIS species.

The abundance studies based on IBEX observations require both careful data analysis and a similarly careful modeling insight. Our paper is part of the latter endeavor. The experimental papers start with \citet{mobius_etal:09b}, where detection of NIS He, Ne, and O in the IBEX signal is reported. A first quantitative look at the Ne/O abundance is presented by \citet{bochsler_etal:12a}. \citet{saul_etal:12a, saul_etal:13a} successfully detected NIS H and traced the evolution of the total flux with ecliptic longitude and from season to season, registering tens of thousands of NIS H atoms and noticing the gradual decay expected due to the solar radiation pressure increasing with the rise of solar activity. Hence, there should not be any problems with the statistics at the NIS H side of the abundance studies.

At the modeling side, we have the present paper and the papers by \citet{sokol_etal:12a} and \citet{bzowski_etal:12b}, where the ionization factors for H and D are discussed. The transmission of NIS D and in some extent of H is discussed in the present paper. 

In a broader astrophysical context the results from the IBEX abundance studies will be a starting point for a broader analysis. This is because they are just point measurements on the spatial scale of the LIC and, as it results from modeling studies, e.g., by \citet{slavin_frisch:08a}, the ionization states of the species in the LIC show gradients due to inhomogeneities in the radiation field and possibly cloud boundaries interaction. The ionization states are of course directly related to the abundances of neutral populations in the local interstellar gas. With all this we believe it is clear that abundance studies by IBEX are complementary to the absorption studies and certainly cannot replace them. On the other hand, they may provide valuable insight into processes operating in interstellar clouds in microscale. 

\section{Simulations}

Two populations of NIS~D are expected in the heliosphere \citep{baranov_etal:91}: the primary flow, coming directly from the LIC, and the secondary component, which is created in the outer heliosheath due to charge exchange between NIS~H and D atoms and interstellar D$^+$ ions in the heated and compressed plasma in front of the heliopause (HP). In this paper we assume that the flow parameters of the two populations inside the HP are homogeneous as a function of the offset angle from the nose of HP and identical with those of the corresponding H populations. We adopt them based on the analysis of pickup ion measurements \citep{bzowski_etal:08a} and kinetic global heliosphere simulations \citep{izmodenov_etal:03b}. Since a recent analysis of the IBEX NIS~He observations \citep{bzowski_etal:12a, mobius_etal:12a, mccomas_etal:12b} resulted in the flow parameters that differ from the previous values \citep{witte:04, mobius_etal:04a}, we also include simulations with these new parameters. The densities of the D populations at the HP were adopted as 15.6~ppm,  distributed between the primary and secondary populations proportionally to the densities of these populations at TS \citep{tarnopolski_bzowski:08a}. For NIS~He we adopted the density determined by \citet{witte:04}. See Table~\ref{tab:isParams} for numerical values.

The directions of inflow of the primary and secondary populations of both H and D were adopted as identical and equal to the inflow direction of NIS He, even though the inflow direction of NIS He and NIS H were determined to differ by $\sim 6^{\circ}$ \citep{lallement_etal:10a}, most probably due to the action of interstellar magnetic field, which modifies the flow of interstellar plasma in the outer heliosheath. 

We decided to neglect this effect in our simulations because its influence on the modeled NIS H and D flux is expected to be minimal. \citet{izmodenov_etal:05a} predict a very weak sensitivity of the filtration of NIS H through the heliospheric interface on the inflow parameters. Furthermore, the uncertainties of the actual inflow direction of NIS gas, of the variation of the flow parameters of the NIS gas as a function of the offset angle from the upwind direction in the outer heliosheath \citep{katushkina_izmodenov:10}, and detailed behavior of the secondary populations in the outer heliosheath in the presence of interstellar magnetic field are poorly known. Therefore, in our opinion, an attempt to take them into account in the present modeling would not necessarily improve the realism of the results. 

For comparison with observed IBEX count rates and to assess the expected foreground, we simulated NIS~H, He, and O. The parameters used in the simulations are compiled in Table~\ref{tab:isParams}.

The cross sections for photoionization, charge exchange, and electron impact ionization of D are practically identical with H, thus leading to identical H and D ionization rates. Important is the difference in radiation pressure acting on H and D due to the differences in atomic mass and Lyman-$\alpha$ resonance frequencies \citep[see][]{tarnopolski_bzowski:08a}. We use the neutral interstellar gas model by \citet{tarnopolski_bzowski:08a, tarnopolski_bzowski:09} (the Warsaw Test Particle Model), modified for the observation geometry, as specified for NIS~He in \citet{bzowski_etal:12a}, as well as the measurement-based radiation pressure and ionization rate. 

The model is fully time- and heliolatitude dependent. Radiation pressure is a function of time and radial velocity of the atom (see Fig.~1 in \citet{tarnopolski_bzowski:08a}), based on observations of the solar Lyman-$\alpha$ line by \citet{lemaire_etal:02, lemaire_etal:05}. For the time series of the total solar Lyman-$\alpha$ flux to calculate the radiation pressure we adopted the composite Lyman-$\alpha$ time series data product from LASP, available through the LISIRD Web page at http://lasp.colorado.edu/lisird/lya \citep{woods_etal:00}. It is shown in Fig.~\ref{fig:jon}, along with the total ionization rate at the equator that includes charge exchange and photoionization \citep{sokol_etal:12a, bzowski_etal:12b}. The solar wind speed and density as a function of heliolatitude for the calculation of charge exchange rates was adopted after \citet{sokol_etal:12a}.

The simulations were performed using the method described in detail by \citet{bzowski_etal:12a} for NIS He (see their Section 2). The calculations were done for three IBEX observation seasons (2009, 2010, and 2011), with the IBEX-Lo field of view taken according to the measured spin axis orientation \citep{hlond_etal:12a}. The simulated flux in the IBEX reference frame was convolved with the collimator transmission function (see Fig.~2 and 3 and Eq.~(1) in \citet{bzowski_etal:12a}) and averaged over the actual time interval of each orbit. The expected count rates were integrated either over the full energy range (i.e., from 0 to the maximum flow energy) or individually over the three lowest IBEX-Lo energy steps. The results correspond to the collimator-averaged flux of NIS atoms in the IBEX-inertial reference frame, mapped as a function of IBEX spin angle. From these, maximum values were taken for each orbit. They are shown in Fig.~\ref{fig:3seasons} for the years after IBEX launch as the maximum flux for a given orbit, as a function of orbit number, which increases with the ecliptic longitude of the observer. Solar activity during this interval was increasing from the minimum conditions. 

We also performed the simulations for the conditions just after solar maximum by shifting the observation time, geometry conditions etc. to the beginning of 2002 (when the solar activity started to decrease), using the respective total ionization rate and radiation pressure history from \citet{sokol_etal:12a} and \citet{bzowski_etal:12b}. The results are also shown in Fig.~\ref{fig:3seasons}. Throughout the paper we use the solar minimum term for the solar activity conditions after IBEX launch and solar maximum for the conditions seven years before the observations. Note that such a treatment of the solar maximum conditions in the context of future IBEX observations should be regarded as conservative because the present solar activity cycle is predicted at approximately half the strength of the previous maximum (\citet{hathaway:09a}; see also solarscience.msfc.nasa.gov/predict.shtml), which should result in a weaker drop of the NIS flux than we predict in the model presented here.

The flux of NIS atoms at a given moment of time is Gaussian-like in spin angle \citep[see, e.g.,][ Fig. 6 through 8]{bzowski_etal:12a}, so the peak height for a given orbit is proportional to the total number of counts for this orbit and enables a quick assessment of the relations between the expected count numbers of different species and their changes with solar activity. To obtain the total count number for a given orbit, one has to integrate the flux over the spin angle and follow the equations specified in the next section.

\section{Results}

\begin{figure}
\resizebox{\hsize}{!}
{\includegraphics[angle=-90]{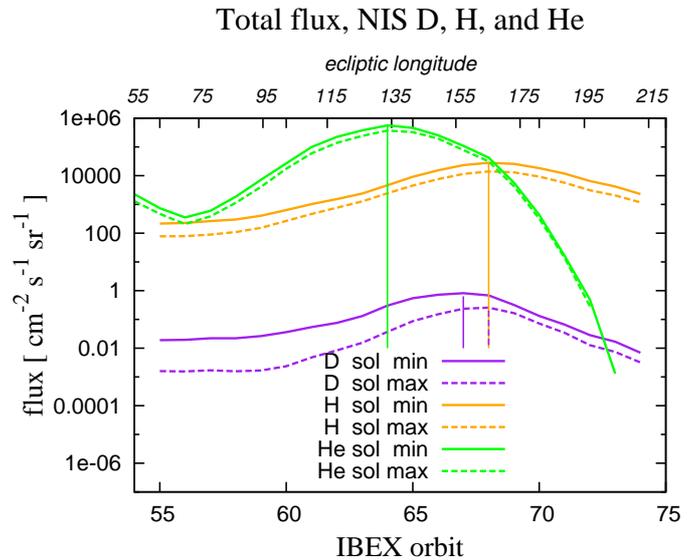}}
\caption{Simulated peak flux for NIS~D and H (primary + secondary populations) and for the primary NIS~He population integrated over the full energy range, taken for the actual observation conditions in 2010 (solid lines) and for solar maximum conditions in 2003 (dashed lines). }
\label{fig:chanHDHe}
\end{figure}

The simulated variation of the flux maximum with ecliptic longitude of the Earth (or IBEX orbit) in Fig. \ref{fig:chanHDHe} confirms that the absolute flux maxima for He, D, and H should be observed in sequence, from negligible radiation pressure (He) to the strongest effect (H). The highest flux values are found for He as expected; the flux of H is lower by 3 orders of magnitude during the orbit of the maximum NIS He flux \citep{mobius_etal:12a, saul_etal:12a}, and the expected D flux is lower than H by 4 orders of magnitude (with somewhat higher abundance at 1~AU than in the LIC due to lower effective radiation pressure acting on D relative to H, as explained by \citet{tarnopolski_bzowski:08a}). The ecliptic longitude at which the flux maximum for a given species occurs does not considerably vary with the interstellar gas speed, inflow longitude, and temperature combinations at the HP that are adopted here, but they depend on the phase in the solar cycle when the observations are made (except for He, for which radiation pressure is zero).

Our simulations confirm the conclusion of \citet{tarnopolski_bzowski:08a} that the kinetic energy of NIS~D in the IBEX frame exceeds the sensitivity threshold only during the Spring observation season. During this interval, the energy of both the primary and secondary NIS~D population falls into the range of energy step 3. At this energy step, the expected D flux is 4-5 orders of magnitude lower than that of NIS~H (the same as in Fig.~\ref{fig:chanHDHe} for the total flux integrated over the full energy range).  

It is expected that the peak fluxes remain stable during solar minimum and slowly decrease with rising solar activity (Fig.~\ref{fig:3seasons}), mostly because of increased solar Lyman-$\alpha$ flux, responsible for radiation pressure (Fig.~\ref{fig:jon}). Conversely, a gradual flux increase is expected for declining solar activity, with a solar cycle amplitude of a factor of $\sim 5$ in the primary and $\sim 10$ in the secondary NIS~D flux, as illustrated in Fig.~\ref{fig:3seasons}. 

\begin{figure}[h]
\begin{tabular}{cc}
\resizebox{\hsize}{!}{\includegraphics[angle=-90]{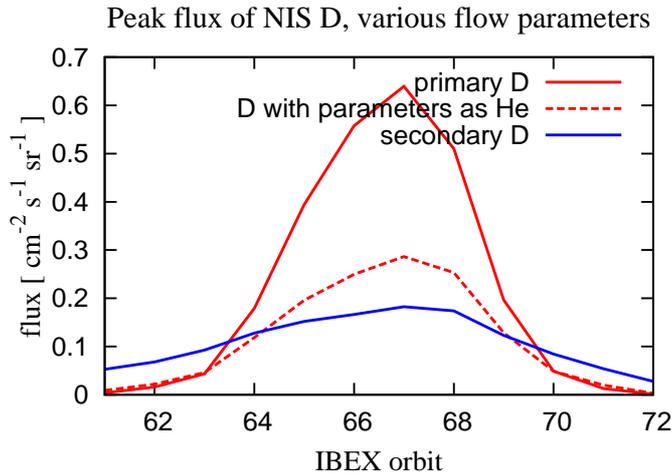}}\\
\end{tabular}
\caption{Simulated peaks of NIS~D flux as a function of IBEX orbit number for the flow parameters as specified in Table~\ref{tab:isParams}: primary component with the He parameters of \citet{witte:04} (solid red), secondary component (blue), and the primary component with the He parameters from the recent analysis by \citet{bzowski_etal:12a} (dashed red). }
\label{fig:popul}
\end{figure}

Like for the other species, the D peak flux at 1~AU varies characteristically along the Earth orbit, reflecting flow velocity and temperature, as illustrated in Fig.~\ref{fig:popul}. With a detector of sufficient sensitivity the flow parameters of the primary NIS~D population could be determined.

However, along with NIS~D, IBEX-Lo will also detect D$^-$ ions sputtered off from the conversion surface by the plentiful NIS~He atoms. To compare the expected counts of sputtered D with those from NIS~D, we integrate the flux of both components in energy step 3 over spin phase and convert the spin-integrated flux into counts for each orbit. 

To calculate the counts $c_{\mathrm{X,n}}$ observed by IBEX-Lo from the spin-integrated flux $F_{\mathrm{X,n}}$ for a species X and energy step $n$, we must fold in the species and energy dependent geometric factor $g_{\mathrm{X,n}}$ of IBEX-Lo, the exposure time $t_{\mathrm{eff}}$, and the duty cycle $q$:
\begin{equation}
\label{eq:counts}
c_{\mathrm{X,n}} = F_{\mathrm{X,n}}\,g_{\mathrm{X,n}}\,q\,t_{\mathrm{eff}}
\end{equation}
Throughout the following calculations, we adopt identical exposure times and duty cycles for all species and all orbits: $t_{\mathrm{eff}}=6$~days and $q = 1/8$ (because of the IBEX-Lo stepping cycle with 8 energy steps). The geometric factor for NIS~D for energy step 3 is adopted identical as for H from IBEX-Lo calibrations: $g_{\mathrm{H,3}} = g_{\mathrm{D,3}} = 2.172\times 10^{-5}$~cm$^2$ sr keV/keV, and for O,  the energy step 6 $g_{\mathrm{O,6}} = 6.888\times 10^{-5}$~cm$^2$ sr keV/keV. 

To calculate the D counts $c_{\mathrm{D,sputt}}$ sputtered from the water layer by interstellar He the energy-integrated flux of He $F_{\mathrm{He,all}}$ must be used. The geometric factor based on the D sputtering efficiency from the terrestrial water layer $g_{\mathrm{sputtD,3}} = 2.483\times10^{-5}$~cm$^2$ sr keV/keV, which is adopted as the measured H sputtering efficiency, and the D abundance in terrestrial water $\xi_{\mathrm{terr}}$ are used, modifying Eq.~(\ref{eq:counts}) to:
\begin{equation}
\label{eq:sputtCounts}
c_{\mathrm{D,sputt}} = F_{\mathrm{He,all}}\,\xi_{\mathrm{terr}}\, g_{\mathrm{sputtD,3}}\,q\,t_{\mathrm{eff}}
\end{equation} 

\begin{figure}
\resizebox{\hsize}{!}
{\includegraphics[angle=-90]{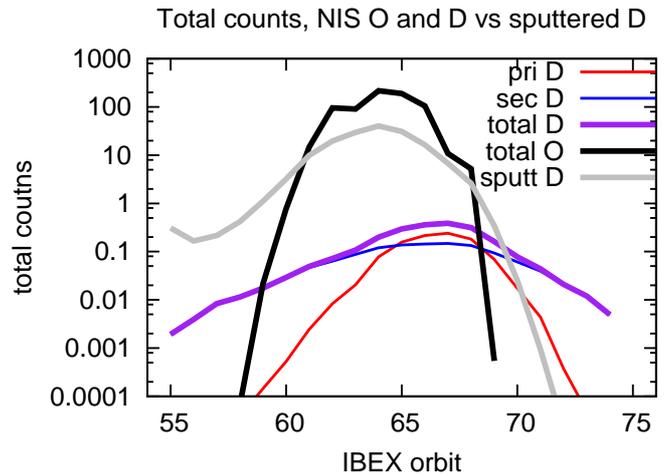}}
\caption{Simulated spin phase-integrated total counts of NIS~D per orbit, compared with the total counts of D atoms sputtered off by NIS~He atoms from the terrestrial water layer on the IBEX-Lo conversion surface (grey), and with the total counts of the primary NIS~O atom population (black). The total counts of D (purple) are a sum of the counts from the primary (red) and secondary populations (blue). Shown are simulations for the 2010 observation season conditions. An exploded view of the interval where the NIS D signal is on a comparable level to the sputtered signal is shown in Fig.~\ref{fig:de_totCounts_D}}
\label{fig:deTotCounts}
\end{figure}

Results of these simulations for the 2010 observation season are shown in Fig.~\ref{fig:deTotCounts}, where the primary and secondary components of NIS~D and their sum are compared with the expected sputtering contribution of D from the water layer. Also shown are the counts of NIS~O, which provide a test of the expected absolute count levels: the model counts are in reasonable agreement with actual observations reported by \citet{bochsler_etal:12a}.

\begin{figure}
\resizebox{\hsize}{!}
{\includegraphics[angle=-90]{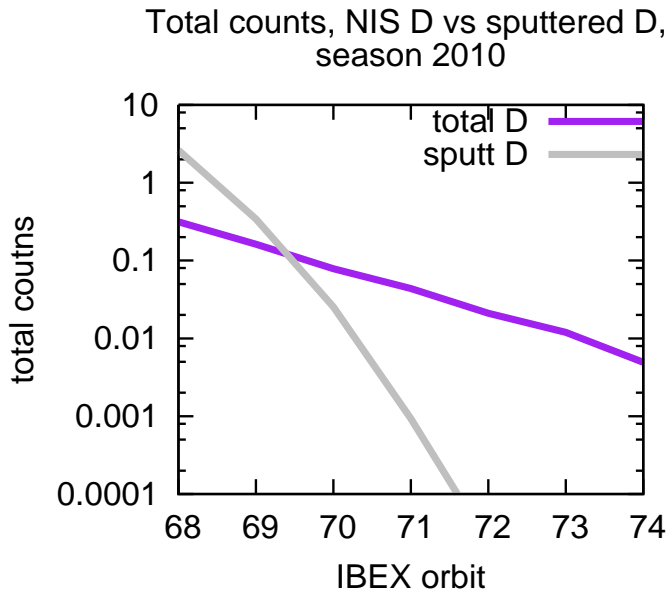}}
\caption{Simulated spin phase-integrated total counts of NIS~D per orbit, compared with the total counts of D atoms sputtered off by NIS~He atoms from the terrestrial water layer on the IBEX-Lo conversion surface (grey) for the interval in 2010 when the counts from the sputtered component are comparable with the counts from NIS D. See Table~\ref{tab:2010}. Note that these atoms almost solely come from the secondary population.}
\label{fig:de_totCounts_D}
\end{figure}

It is clear from the figure that during most of the NIS observation season, the sputtered D signal will dominate over interstellar D. From the whole 2010 season, 160 counts of sputtered D and only 2.2 of NIS D are expected, with similar proportions in 2009 and 2011. However at the beginning of March each year (i.e., starting from Orbit 20 in 2009, Orbit 68 in 2010, and Orbit 116 in 2011), the sputtered and NIS D counts become comparable, with the sputtered component quickly falling off. Thus until March each year, the observed D/H abundance should be stable and close to the terrestrial value. Afterwards, a rapid change is expected. From the simulations we performed, it is expected that the number of NIS D atoms observed on orbits 20 through 26 is $\sim 0.58$, on Orbits 68 through 74 is 0.64, and on Orbits 116 through 122 is 0.38, i.e. total of $\sim 1.6$ during the 3 intervals. Simultaneously, the expected tally of the sputtered D atoms during these intervals is expected at 2.6 in 2009, 3.0 in 2010, and 1.5 in 2011, total of $\sim 7$ during the 3 seasons. See Tables \ref{tab:2009}, \ref{tab:2010}, and \ref{tab:2011} and Fig.~\ref{fig:de_totCounts_D} for details. These numbers were obtained assuming coverage of the observation time close to 100\% for each orbit. This is realistic because no spin phase information is needed, only the total spin phase-integrated counts. Therefore, any spin phase synchronization issues that have so far reduced the effective observation time for other species \citep{mobius_etal:12a} are not important here. 

Thus, extracting a genuine NIS~D signal from the IBEX data will be challenging. It will require a careful selection of the time interval when the NIS D signal is not expected to be swamped in the sputtered D signal,  integrating thus selected signal over all available observation seasons and performing an in-depth statistical analysis of the signal to appropriately account for the remnant sputtered component.

\section{Discussion}

One has to realize the predictions we offer are based on assumptions on absolute densities of NIS He and NIS D at the termination shock that need not be accurate, even though we have adopted the best information currently available. The absolute density of NIS He was taken from \citet{witte:04}. It had been obtained from analysis of NIS He observations by GAS/Ulysses. The flow parameters from GAS/Ulysses have been challenged \citep{bzowski_etal:12a, mobius_etal:12a} based on analysis of observations from IBEX-Lo, but no new estimates of the NIS He density at the termination shock have been provided. The assumed absolute density of NIS D at TS is calculated based on a recent determination of NIS H density at the termination shock based on Ulysses observations of H pickup ions \citep{bzowski_etal:08a}. This, however, required the assumption that the NIS D abundance in front of the heliosphere is equal to the abundance averaged over the Local Cloud, and the assumption that the parameters of the primary and secondary populations of NIS D at the termination shock are identical with the parameters of NIS H.  

\begin{table*}
\caption{Estimated count numbers from the sputtered D component and genuine NIS D atoms shown for subsequent IBEX orbits in 2009}
\label{tab:2009}
\centering
\begin{tabular}{l c c c c c c c c}
\hline\hline
orbit & 20 & 21 & 22 & 23 & 24 & 25 & 26 & total\\
sputtered~D & $1.72$ & $8.02\cdot10^{-1}$ & $2.99\cdot10^{-2}$ & $3.10\cdot10^{-3}$ & $2.75\cdot10^{-5}$ & 0 & 0 & 2.56 \\
NIS~D& $2.39\cdot10^{-1}$ & $1.98\cdot10^{-1}$ & $6.72\cdot10^{-2}$ & $4.37\cdot10^{-2}$ & $2.00\cdot10^{-2}$ & $1.00\cdot10^{-2}$ & $5.55\cdot10^{-3}$ & $0.58$ \\
\hline
\end{tabular}
\end{table*} 

\begin{table*}
\caption{Estimated count numbers from the sputtered D component and genuine NIS D atoms shown for subsequent IBEX orbits in 2010}
\label{tab:2010}
\centering
\begin{tabular}{l c c c c c c c c}
\hline\hline
orbit & 68 & 69 & 70 & 71 & 72 & 73 & 74 & total\\
sputtered~D & 2.62 & $3.42\cdot10^{-1}$ & $2.54\cdot10^{-2}$ & $9.30\cdot10^{-4}$ & $2.28\cdot10^{-5}$ & $2.66\cdot10^{-8}$ & 0 & 2.99 \\
NIS~D& $3.16\cdot10^{-1}$ & $1.62\cdot10^{-1}$ & $7.83\cdot10^{-2}$ & $4.39\cdot10^{-2}$ & $2.10\cdot10^{-2}$ & $1.20\cdot10^{-2}$ & $4.94\cdot10^{-3}$ & $0.64$ \\
\hline
\end{tabular}
\end{table*} 

\begin{table*}
\caption{Estimated count numbers from the sputtered D component and genuine NIS D atoms shown for subsequent IBEX orbits in 2011}
\label{tab:2011}
\centering
\begin{tabular}{l c c c c c c c c}
\hline\hline
orbit & 116 & 117 & 118 & 119 & 120 & 121 & 122 & total\\
sputtered~D & 1.21 & $2.76\cdot10^{-1}$ & $6.72\cdot10^{-4}$ & $1.32\cdot10^{-5}$ & 0 & 0 & 0 & 1.48 \\
NIS~D& $1.87\cdot10^{-1}$ & $1.25\cdot10^{-1}$ & $3.18\cdot10^{-2}$ & $1.78\cdot10^{-2}$ & $1.08\cdot10^{-2}$ & $4.42\cdot10^{-3}$ & 0 & $0.37$ \\
\hline
\end{tabular}
\end{table*} 

These assumptions have not been verified by modeling, but in our opinion seem realistic. There is no observational evidence suggesting that the ionization state of D in the LIC differs from the ionization state of H, given their almost identical ionization potential and the relatively small difference in thermal speeds. The filtration of NIS D through the heliosphere is expected to be very similar to the filtration of H. The secondary population of NIS D (i.e., the population created in the outer heliosheath due to the charge exchange reaction between the outer heliosheath plasma and NIS gas) comes up predominantly from the reaction of interstellar D$^+$ ions with interstellar H atoms, i.e., identically as the secondary population of NIS H. \citet{izmodenov:07} studied the filtration of NIS H through the heliospheric interface and concluded that the filtration factor weakly depends on the interstellar gas parameters. Filtration should deplete the primary D population, just as the primary H, and the flow parameters will be weakly modified, similarly as for NIS H. The only kinematic difference between H and D will be due to the small difference in the thermal speeds. Thus also the secondary population of NIS D will likely have similar flow parameters to the secondary population of NIS H. The exact parameters have not been modeled, but the general relations of the flow parameters of the secondary relative to the primary are expected to be similar: the flow will be considerably decelerated and the temperature increased. We have conservatively assumed an agreement in the temperatures of the secondary H and D, and not in thermal speeds. If the thermal speeds were to be equal, it would effectively heat up the secondary D, which would result in a longer duration of the NIS D signal after the sputtered component vanishes, and effectively would likely increase the chances for detection.

The sputtering efficiency for D from the water layer and of registering the NIS D atoms directly has not been calibrated prior to flight. The values used in the paper are deduced based on calibrated efficiencies for H and He. Resulting uncertainty does not qualitatively change the conclusion that the sputtered component will dominate over the interstellar, it may only modify the break-even point. If the signal from sputtered D is detected, then the sputtering efficiency can be effectively measured from comparison of the sputtered counts versus NIS He counts.

Departures from these assumptions may result in a change both in the absolute values of total counts predicted and in the proportions between the sputtered and NIS signal. The change in proportions between the sputtered and NIS intensities should mostly affect orbits 20 and 21 and their equivalents in the subsequent years because on these orbits the contribution from the sputtered component is strongest. The later orbits should be much less affected because the sputtered signal is predicted to fall off very rapidly. 

In the simulations we assumed that the inflow directions of NIS He and NIS H (both populations) are identical. This is not exactly true, as concluded by \citet{lallement_etal:05a}, who suggest that the longitude of the H direction is lower by $\sim 3\degr$ than of the He direction. The H flow observed by \citet{lallement_etal:05a} is an inseparable superposition of the primary and secondary population. We estimate that this effect should shift the locations of the total D line relative to the sputtered D line in Fig.~\ref{fig:deTotCounts} by a half of the orbit leftward (because one orbit corresponds to approximately 7\degr), which would reduce a little the interval when the NIS D signal is above the sputtered signal. The change would be small, however, and of little significance for the conclusions. 

Taking all this into consideration, the numbers we give should be regarded as order estimates accurate to a factor of $<10$. The general conclusion from our simulation is that the integrated signal from the orbits from March each year should bring a few genuine NIS D atoms and a comparable, a little larger number of counts of the sputtered D atoms. The NIS D should be visible in the IBEX-Lo data as a change in the observed D/H abundance from the terrestrial value, starting from March each year.

An interesting aspect of observations of NIS D, resulting from our study, is that potentially detectable NIS atoms will most likely belong to the secondary population (cf the blue and red lines in Fig.~\ref{fig:deTotCounts}). Thus the ability to build up enough statistics of NIS D atoms is welcome since it provides direct insight into the physical state of the gas in the outer heliosheath.

Direct observation of neutral interstellar D provides a highly complementary local measurement of D at the location of the Sun. Simultaneously, it provides an opportunity to improve our understanding of the interaction of the interstellar gas and plasma with the heliosphere, because with H and D we have an isotope pair with both of them substantially affected by the plasma flow around the heliosphere and by solar radiation pressure. Admittedly, based on the analysis presented here the first observations will contain a large uncertainty because only a few counts are expected over the past 3-4 years of operation. However, this modeling study together with the observational analysis will provide the basis for improved observations over a longer IBEX mission and for improvements of the observational technique on future missions.

Given the low expected count number, there is a $\sim 20$\% Poisson probability for non-detection of NIS D. Another explanation for potential non-detection might be a significantly lower than assumed absolute density of neutral interstellar gas or lower abundance of D in front of the heliosphere. Detection at a level comparable to the prediction would likely suggest that the picture of the heliosphere and the LIC we have is mostly correct. Anyway, it would be a totally independent confirmation of a result obtained with different methods (spectroscopy) and would also show that within large error bars, there are no significant deviations of the D/H abundance from the LIC average. Detection at a level statistically significantly higher than the predictions could imply a higher total density of interstellar matter in front of the heliosphere or a large overestimation of the attenuation of NIS D inside the heliosphere (which we believe unlikely). 

The detection capability for NIS~D could be further improved by increasing the sensor duty cycle by up to a factor of eight, if viewing time is concentrated on energy step 3, n which interstellar D is primarily expected, i.e., by operating in an IBEX-Lo Interstellar Medium flow mode \citep{fuselier_etal:09b} optimized for D. Preferably, such a campaign should be carried out during future low solar activity, when the expected D signal is strongest. Increasing the statistics by a factor of $\sim 10$ would enable making statistical tests of various hypotheses related to the assumptions used in this study. Obtaining the D/ H abundance on a precision level of a few times $10^{-2}$, comparable to the best astrophysical estimates, requires the ability to estimate the counts of both species with a precision level of $10^{-2}$, which appears out of reach for D with the current instrument and so likely has to await a follow-on mission. The counting statistics for H does not seem to be an issue here.

Because interaction of the incoming interstellar gas with the heliospheric boundary and extinction of species will differently affect each individual abundance ratio, observations of the D/H ratio have to be seen in context with other abundance studies based on IBEX data, such as Ne/O \citep{bochsler_etal:12a} and on pickup ion data \citep{geiss_gloeckler:98a, gloeckler_geiss:04a, gloeckler_etal:09a}. It should be noted though that finding the D/H ratio from IBEX observations is probably the best bet over deriving any ratios with other species, such as D/O. For both D and H the process of conversion to a negative ion, which determines the sensor efficiency for these species, is almost identical, while the affinity to an electron for O is much greater. In addition, IBEX-Lo records a mix of Ne and O, where Ne is detected through sputtering on its conversion surface \citep{bochsler_etal:12a}. Therefore, additional systematic errors will have to be considered when reporting D/O ratios that do not affect the D/H ratio and, as suggested by a comparison of expected fluxes in Fig. \ref{fig:deTotCounts}, the modeling part of the analysis would be equally complex as for the D/H pair.

\section{Summary and conclusions}
Based on pre-launch estimates by \citet{tarnopolski_bzowski:08a}, it was expected that IBEX would be able to detect NIS~D. We performed simulations of the expected NIS~D signal based on actual interstellar flow measurements for times, geometry, and heliospheric conditions during minimum and maximum of solar activity using a state of the art kinetic model of the NIS gas flow in the inner heliosphere. We also compared this signal with the expected foreground of D sputtered by NIS~He from the terrestrial water layer that covers the IBEX-Lo conversion surface. 

The entire NIS~D signal is expected within the energy range of IBEX-Lo energy step 3. The expected NIS~D maximum occurs two IBEX orbits after the peak of the He signal and approximately one orbit before the peak of H. The expected D/H ratio in energy step 3 is higher during solar maximum than during solar minimum, while the fluxes of both species are reduced during solar maximum due to the increased radiation pressure. 

As an additional challenge, a foreground of NIS~D generated from NIS~He due to sputtering from terrestrial water on the IBEX-Lo conversion surface exceeds the expected NIS~D signal by almost two orders of magnitude. Focusing on IBEX orbits after the sputtered signal and integrating the observations over spin phase will maximize the detection capability. The expected count numbers from the sputtered and NIS populations become comparable after the beginning of March each year and can hopefully be interpreted using appropriate statistical methods. During this time, we expect $\sim 2$ counts of NIS D collectively in 2009, 2010, and 2011 and a few counts from the sputtered component during this time interval. In fact, the interval when the sputtered population stops overwhelming the interstellar can be found from observations: the observed D/H abundance in the signal should be stable and close to terrestrial during the times when mostly the sputtered component is present. Afterwards, the observed D/H ratio should change because the contribution from the NIS component should become statistically significant. 

Increasing the duty cycle by concentrating observations on IBEX-Lo energy step 3 in a dedicated NIS~D observation mode for several seasons during future solar minimum conditions would further boost the NIS D detection. Under these conditions, undoubtful detection of NIS~D with IBEX appears potentially possible. Fortunately, IBEX was recently maneuvered into a previously unknown, long-term stable lunar synchronous orbit \citep{mccomas_etal:11a} with a significantly reduced radiation environment. Given an excellent state-of-health and hydrazine reserves of the spacecraft, IBEX could function for more than a full solar cycle, possibly for decades, potentially providing opportunity to build up the necessary statistics. 

Our estimates also provide guidance for future missions, such as the Interstellar Mapping (IMAP) mission suggested by \citet{McComas_etal:11b}, which would carry next generation neutral atom cameras, with reduced internal D contamination and substantially improved collection power.

\begin{acknowledgements}
M.A.K., M.B. and J.M.S. were supported by the Polish Ministry for Science and Higher Education grant N-N203-513-038, managed by the Polish National Science Center. M.A.K was stipendiary of the START program of the Polish Science Foundation for 2012. E.M. and D.J.M. were supported by the IBEX project under contract NNG05EC85C. E.M. was also supported under grant NNX10AC44G. The SRC PAS and UniBern members of the author team gratefully acknowledge the hospitality of the International Space Science Institute (ISSI) in Bern, Switzerland.
\end{acknowledgements} 

\bibliographystyle{aa}
\bibliography{iplbib}

\end{document}